\documentstyle[preprint,prd,aps,tighten,epsf]{revtex}
\begin{document}
\draft
\newcommand{\nl}{\nonumber \\}
\newcommand{\bea}{\begin{eqnarray}}
\newcommand{\eea}{\end{eqnarray}}
\newcommand{\bi}{\bibitem}
\newcommand{\be}{\begin{equation}}
\newcommand{\ee}{\end{equation}}
\newcommand{\bt}{\begin{table}}
\newcommand{\et}{\end{table}}
\newcommand{\btab}{\begin{tabular}}
\newcommand{\etab}{\end{tabular}}
\def\eg{{\it e.g.}}
\def\ra{\rightarrow}
\def\dek#1{\times10^{#1}}
\def\piee{K^+\ra\pi^+e^+e^-}
\def\pimm{K^+\ra\pi^+\mu^+\mu^-}
\def\pill{K^+\ra\pi^+\ell^+\ell^-}
\def\die{e^+e^-}
\def\dim{\mu^+\mu^-}
\def\dil{\ell^+\ell^-}
\def\kmu{K^+\ra\mu^+\nu_\mu}
\def\ratio{\dim/\die}
\def\hh#1#2#3{$#1^{+#2}_{-#3}$}
\def\ffpi{F_{\pi^+}(t)}
\def\ffkp{F_{K^+}(t)}
\def\ffk0{F_{K^0}(t)}
\def\fpi{f_{\pi^+}}
\def\fkp{f_{K^+}}

% Journal and other miscellaneous abbreviations for references
\def \zpc#1#2#3{Z.~Phys.~C {\bf#1}, #2 (19#3)}
\def \plb#1#2#3{Phys.~Lett.~B {\bf#1}, #2 (19#3)}
\def \prl#1#2#3{Phys.~Rev.~Lett.~{\bf #1}, #2 (19#3)}
\def \pr#1#2#3{Phys.~Rep.~{\bf #1}, #2 (19#3)}
\def \prd#1#2#3{Phys.~Rev.~D~{\bf#1}, #2 (19#3)}
\def \npb#1#2#3{Nucl.~Phys.~B {\bf#1}, #2 (19#3)}
\def \rmp#1#2#3{Rev.~Mod.~Phys.~{\bf#1}, #2 (19#3)}
\def \ea{{\it et al.}}
\def \ibid#1#2#3{{\it ibid.} {\bf#1}, #2 (19#3)}
\def \sjnp#1#2#3#4{Yad. Fiz. {\bf#1}, #2 (19#3) [Sov. J. Nucl. Phys.
{\bf#1}, #4 (19#3)]}
\def \pan#1#2#3#4{Yad. Fiz. {\bf#1}, #2 (19#3) [Physics of Atomic Nuclei
{\bf#1}, #4 (19#3)]}
\def \epjc#1#2#3{Eur. Phys. J. C {\bf#1}, #2 (19#3)}
\def \appb#1#2#3{Acta Phys. Pol. B {\bf#1}, #2 (19#3)}

\title{\bf 
About one long-range contribution to $\pill$ decays
}
\author{Peter Lichard\footnote{On leave of absence from Department of 
Theoretical Physics, Comenius University,
842-15 Bratislava, Slovak Republic.}}
\address{
Department of Physics and Astronomy, University of Pittsburgh,
Pittsburgh, Pennsylvania 15260; \\
and Institute of Physics, Silesian University, 746-01 Opava, Czech Republic
}
\maketitle
\begin{abstract}
We investigate the mechanism of $\pill$ ($\ell=e$,~$\mu$) decays in which
a virtual photon is emitted either from the incoming $K^+$ or the outgoing
$\pi^+$. We point out some inconsistencies with and between two
previous calculations, discuss the possible experimental inputs, and 
estimate the branching fractions. This mechanism alone fails to
explain the existing experimental data by more than one order-of-magnitude. 
But it may show itself by its interference with the leading long-range
mechanism dominated by the $a_1^+$ and $\rho^0$ mesons.
\end{abstract}

\pacs{PACS number(s): 12.15. y, 12.15.Ji, 13.20.Eb, 13.40.Gp}
% 12.15.y Electroweak interactions
% 12.15.Ji Application of electroweak models to specific processes
% 13.20.Eb Decays of kaons
% 13.40.Gp Electromagnetic form factors

\narrowtext

With the new data on the $\pill$ decays coming soon \cite{hong}, thoughts 
about the possible mechanism(s) behind them are gaining steam. One candidate
is virtual photon emission from the incoming $K^+$ and outgoing $\pi^+$; 
it was conjectured recently that its contribution to the decay amplitude 
may be important \cite{jim}. The purpose of this note is to scrutinize 
the role of this mechanism.

The matrix element of this mechanism was given in a paper by Vainshtein, 
Zakharov, Okun, and Shifman \cite{vain} and used in the same form 
by Bergstr\"{o}m and Singer \cite{berg}. After changing to contemporary 
notations and to the $(+,-,-,-)$ metrics, the matrix element becomes
\bea
{\cal M}&=&\frac{e^2G_F}{\sqrt{2}}(\fpi V_{ud})(\fkp V^*_{us})
\frac{1}{m_K^2-m_\pi^2}
\left[\frac{m_K^2}{6}\langle r^2\rangle_{K^+}-
\frac{m_\pi^2}{6}\langle r^2\rangle_{\pi^+}\right] \nl
&\times&
\left( p_K +p_\pi\right)^\mu\ 
\overline{u}(p_-)\gamma_\mu v(p_+)\ .
\label{vainformula}
\eea
The formula we derive in this note differs from (\ref{vainformula}). Another 
formula exists in the literature \cite{eilam}, which differs from ours ``only"
in the multiplicative constant. Despite the fact that our formula
differs more substantially from Eq.~(\ref{vainformula}) than it does from that
in \cite{eilam}, we consider that our formula is more closely related
to Eq.~(\ref{vainformula}) than to the latter in the underlying physics, 
as we will discuss later.

Because of these discrepancies we present the derivation of our formula in  
thorough detail. But let us first define, for completeness 
and later reference, the quantities which enter Eq.~(\ref{vainformula}). 
$G_F$ is the Fermi constant related to the weak coupling constant, $g$, 
and to the $W^\pm$ boson mass by $g^2=4\sqrt{2}G_Fm_W^2$. The decay constant 
of the $\pi^+$ meson is defined by
\[
\langle 0|\bar{d}(0)\gamma^\mu\gamma_5u(0)|p\rangle_{\pi^+}=
i\ f_{\pi^+}\ p^\mu\ .
\]
The definition of the $K^+$ decay constant, $f_{K^+}$, is analogous. The $V$s
are the elements of the Cabibbo-Kobayashi-Maskawa matrix. Using the $K^+$
and $\pi^+$ mean lifetimes, the branching
fractions of  decays $\pi^+\ra\mu^+\nu_\mu$ and $K^+\ra\mu^+\nu_\mu$
given in \cite{pdg}, and the method of radiative corrections
described there, we obtained the following values:
\bea
\left|f_{\pi^+}V_{ud}\right|^2 &=& (1.6418\pm0.0010)
\dek{-2}\ {\rm GeV}^2,\nl
\left|f_{K^+}V_{us}\right|^2 &=& (1.2471\pm0.0044)
\dek{-3}\ {\rm GeV}^2 ,\nonumber
\eea
which we use below in numerical calculations. The mean square radii (MSR) 
are defined by means of the low $|t|$ expansion of the electromagnetic 
form factor
\be
F(t)=F(0)+\frac{\langle r^2\rangle}{6}\ t+\cdots\ ,
\label{lowt}
\ee
where $F(0)=1$~(0) for charged (neutral) mesons. Finally, spinors
$u$ and $v$ in Eq.~(\ref{vainformula}) refer to the outgoing leptons.

Let us now derive our formula.
When the electromagnetic interaction is switched off, the dynamics of the 
system containing charged pions, kaons, and weak-gauge bosons is described 
by the effective Lagrangian\footnote{In \cite{md} we used a simpler
effective Lagrangian with a contact weak interaction between mesons
to show that, for the point-like mesons, the matrix element considered
here vanishes. We checked that the simpler Lagrangian gives the 
same results as presented here, but prefer to be more rigorous.}
\be
{\cal L}_0={\cal L}_{\protect{\rm free}}+{\cal L}_{\protect{\rm weak}},
\label{l0}
\ee
where ${\cal L}_{\protect{\rm free}}$ is the well-known Lagrangian of
the free $\pi^\pm$, $K^\pm$, and $W^\pm$ fields, and
\be
{\cal L}_{\protect{\rm weak}}=-i\frac{g}{2\sqrt{2}}W^\dagger_\mu
\left(f_{\pi^+}V_{ud}^*\partial^\mu\phi_\pi+
f_{K^+}V_{us}^*\partial^\mu\phi_K\right)+ {\rm H.c.}
\label{weaklag}
\ee
After turning on the electromagnetic interaction by the minimal
substitution 
$\partial^\mu\ra\partial^\mu +ieA^\mu$, the Lagrangian
acquires additional terms and becomes
\[
{\cal L}={\cal L}_0+{\cal L}_\gamma+{\cal L}_{\gamma\gamma}\ .
\]
The last part contains the product of two electromagnetic field
operators and does not operate in our case. The one-photon part
contains two terms
\[
{\cal L}_\gamma=
{\cal L}_{\gamma,{\rm free}}+{\cal L}_{\gamma,{\rm weak}}.
\]
The former was induced from the free part of the Lagrangian (\ref{l0}), 
and the latter from its weak interaction part (\ref{weaklag}). 
They are given by
\bea
{\cal L}_{\gamma,{\rm free}}&=&  ieA^\mu\sum_{P=\pi,K}
\left[\left(\partial_\mu
\phi_P^\dagger\right)\phi_P-\phi_P^\dagger\partial_\mu\phi_P
\right]\ ,\nl
{\cal L}_{\gamma,{\rm weak}}&=& \frac{eg}{2\sqrt{2}}A^\mu W^\dagger_\mu
\left(f_{\pi^+}V_{ud}^*\phi_\pi+
f_{K^+}V_{us}^*\phi_K\right)+ {\rm H.c.} \nonumber
\eea
The amplitude of the (virtual) photon emission from $W^\pm$ is suppressed 
by a factor of $(m_K/m_W)^2$, and hence, the corresponding term in 
${\cal L}_{\gamma,{\rm free}}$ is not shown.

To proceed further, we need to determine the interaction
Hamiltonian. We can show that the relation
\[
{\cal H}_{\rm int.}=-{\cal L}_{\rm weak}-{\cal L}_\gamma
-{\cal L}_{\gamma\gamma}
\]
holds, up to non-covariant terms which vanish when the matrix element
is calculated between the physical states \cite{nishijima,itzykson}.
After deriving the Feynman rules in momentum space, we modified them by
multiplying the electric charge of mesons by the corresponding form factors
$F_P(q^2)$ ($P=\pi^+$, $K^+$), thereby accounting for the $\pi^+$ and 
$K^+$ internal structure. As a result, we obtain the following 
vertex (junction) factors, starting with the well-known electromagnetic
vertex induced from the free Lagrangian
\begin{itemize}
\item
$-ieF_P(q^2)(p_a+p_b)^\mu$ for the $P\ra P\gamma$
vertex. Here, $p_a$ ($p_b$) is the four-momentum of the incoming (outgoing)
meson, $q=p_a-p_b$ is the (virtual) photon momentum, and $\mu$ is the
index connected with the photon line;
\item
$-ig/(2\sqrt{2})\ f_{K^+}V_{us}^*\ p^\mu$ for the $K^+\ra W^+$ junction;
\item
$-ig/(2\sqrt{2})\ f_{\pi^+}V_{ud}\ p^\mu$ for the $W^+\ra\pi^+$ junction;
\item
$ieg/(2\sqrt{2})\ F_{K^+}(q^2)\ f_{K^+}V_{us}^*\ g^{\mu\nu}$
for the $K^+\ra W^+\gamma$ vertex;
\item
$ieg/(2\sqrt{2})\ F_{\pi^+}(q^2)\ f_{\pi^+}V_{ud}\ g^{\mu\nu}$
for the $W^+\ra\pi^+\gamma$ vertex.
\end{itemize} 
The contributions to the matrix element of the $\pill$ decay are
depicted in Fig.~1. Their evaluation leads to
\bea
{\cal M}&=&\frac{e^2G_F}{2\sqrt{2}}(\fpi V_{ud})(\fkp V^*_{us})\
\frac{m_K^2+m_\pi^2}{m_K^2-m_\pi^2}\ \frac{\ffkp-\ffpi}{t} \nl
&\times&
\left( p_K +p_\pi\right)^\mu\ 
\overline{u}(p_-)\gamma_\mu v(p_+)\ ,
\label{myme}
\eea
where $t=(p_K-p_\pi)^2$. When writing (\ref{myme}) we took advantage
of the fact that 
the contraction of $(p_K-p_\pi)^\mu$ with the lepton term vanishes.
Finally, if we ignore possible contributions to the matrix
element from other mechanisms, we  get
the following formula for the differential decay rate in the 
$\dil$ mass $M=\sqrt{t}$\ :
\bea
\frac{d\Gamma_{K^+\ra\pi^+\dil}}{dM}&=&\frac{G_F^2\alpha^2}{48\pi m_K^3}
\left|f_{\pi^+}V_{ud}\right|^2\left|f_{K^+}V_{us}\right|^2
\left(\frac{m_K^2+m_\pi^2}{m_K^2-m_\pi^2}\right)^2 \nl
&\times& \lambda^{3/2}(m_K^2,m_\pi^2,t)\
\sqrt{t-4m_\ell^2}\left(1+\frac{2m_\ell^2}{t}\right)
\left|\frac{F_{K^+}(t)-F_{\pi^+}(t)}{t}\right|^2\ ,
\label{kpiee}
\eea
where $\lambda(x,y,z)=x^2+y^2+z^2-2xy-2xz-2yz$.
All quantities on the left-hand side of Eq.~(\ref{kpiee}) are well
known, except the form factors. We need their values in the time-like 
region below the physical thresholds, where they are inaccessible 
to direct measurement. Only for $F_{\pi^+}(t)$ a small part 
\be
4m_\pi^2<t<(m_K-m_\pi)^2
\label{cut}
\ee
of the $\pill$ $t$-interval lies in the physical region
of the $\die$ annihilation. To determine the form factors in
the region of our interest, one has to combine the experimental
information about them in both the space-like and time-like
regions with their analytic properties. This was undertaken, \eg, in
Ref.~\cite{dubn}. Unfortunately, in the small-$t$ region, both the $\pi^+$
and $K^+$ form factors differ only a little from  unity, which greatly
increases the relative error of their difference. Therefore, one must seek
alternative methods for determining the difference in form factors
in (\ref{kpiee}); we suggest a few below. Two are based on the low 
$|t|$ expansion of the electromagnetic form factor (\ref{lowt}). Most of 
our methods rely on the vector meson dominance (VMD) description of the 
form factors of pions and kaons. So, first, we sketch the necessary 
VMD formulas.

In VMD, the form factors are described by assuming that the pseudoscalar
mesons couple to the photon via the vector meson resonances $\rho^0$,
$\omega$, and $\phi$. Let the functions $V(t)$ ($V=\rho$, $\omega$, 
$\phi$), normalized by the condition $V(0)=1$, describe the contributions
of individual resonances to the pseudoscalar meson form factors.
The normalization conditions imposed on the form factors, together
with the assumptions about the isospin invariance of the strong
vertices, lead to the formulas
\bea
F_{\pi^+}(t)&=&\rho(t)\ , \nl
F_{K^+}(t)&=&\ \frac{1}{2}\rho(t)+c\ \omega(t)+
\left(\frac{1}{2}-c\right)\phi(t)\ ,\nl
F_{K^0}(t)&=&-\frac{1}{2}\rho(t)+c\ \omega(t)+
\label{vmdk0}
\left(\frac{1}{2}-c\right)\phi(t)\ ,\\
\label{fk0asdiff}
\ffk0&=&\ffkp-\ffpi\ .
\eea
In principle, the constant $c$ can be fixed by assuming the (flavor)
SU(3) invariance. Because the latter is violated, we will, instead,
consider $c$ a phenomenological parameter, and fix its value by experimental
data. But before proceeding further, we must specify 
the functions $V(t)$. For narrow resonances $\omega$ and $\phi$,
which, in addition, do not have any open decay channels in our
$t$-range, we can use the prescription 
\be
V(t)=\frac{m_V^2}{m_V^2-t}\ ,
\label{simplev}
\ee
which originates from the free-vector-particle propagator, taking
into account that the term in its numerator containing four-momenta 
does not contribute. 

But the situation with the $\rho^0$ is more complicated. This is evident
from the fact that using the simple formula (\ref{simplev})
for $\rho(t)$ gives the MSR of $\pi^+$ equal to 
$(0.3940\pm0.0008)$~fm$^2$, which disagrees with the experimental findings
that give $(0.439\pm0.008)$~fm$^2$ \cite{amen1}. 
We will, therefore, use the form that properly accounts for
the dynamics of resonances (see, \eg, \cite{ratio})
\be
\label{running}
\rho(t)=\frac{m_\rho^2(0)}{m_\rho^2(t)-t-im_\rho\Gamma_\rho (t)}\ ,
\ee
where $m_\rho^2(t)$ is the running mass squared, $m_\rho$ is the
resonant mass, and $\Gamma_\rho (t)$ 
is the variable total width. The latter vanishes in our kinematic 
range, with the exclusion of the small region already mentioned (\ref{cut}). 
We take the function $m^2_\rho(t)$ from our recent work \cite{ratio};
this gives the $\pi^+$ MSR of $(0.446\pm0.006)$~fm$^2$, 
in good agreement with data. 

Below, we briefly describe six methods for evaluating the form-factor 
difference, which we need to be able to calculate
the branching fraction (\ref{kpiee}) of the $\pill$ decays. The 
corresponding results for both the $\die$ and $\dim$ modes are shown in 
Table~1.
\begin{enumerate}
\item
Both $\ffpi$ and $\ffkp$ are taken from Ref.~\cite{dubn}. We do
not have any way to assess the errors, so the results are
shown with the ``approximate" sign.
\item
Formula (\ref{fk0asdiff}) is used, together with the parametrization
of $\ffk0$ from \cite{dubn}. The same caveat as above applies.
\item
Low-$t$ expansion (\ref{lowt}) is used with 
\be
\label{mrsdiff}
\langle r^2\rangle_{\pi^+}-\langle r^2\rangle_{K^+}=
(0.100\pm0.045)~{\rm fm}^2\ ,
\ee
taken from Ref. \cite{amen2}.
\item
The formula (\ref{fk0asdiff}) and parametrization (\ref{lowt}) for 
$\ffk0$ are used with
\be
\label{mrsk0}
\langle r^2\rangle_{K^0}=-(0.054\pm0.026)~{\rm fm}^2.
\ee
This value was determined in \cite{molzon} by measuring the coherent 
regeneration of $K^0_S$s by atomic electrons; the method was proposed
by Zel'dovich \cite{zeldovich}.
\item
Formulas (\ref{vmdk0}) and (\ref{fk0asdiff}) are used with $c=0.07\pm0.29$,
which was determined by matching the MSR difference (\ref{mrsdiff}).
\item
As above, but the MSR of $K^0$ (\ref{mrsk0}) was matched with
$c=0.36\pm0.17$.
\end{enumerate}    
Of all of those methods, we believe that the last two are the most reliable. 
As we argued, the calculation based directly on the form factors (method 1)
is hampered by the large relative errors in their difference. Method 2, 
using the $K^0$ form factor, is probably more reliable, but similarly here, 
we cannot estimate the final error. Methods 3 and 4 lead to a constant form 
factor of the $\pill$ decay, defined below in Eq.~(\ref{ffdef}), which is 
a bad approximation \cite{alliegro}. Our preference for methods 5 and 6 stems 
from the fact that the systematic errors of the VMD method are certainly 
smaller than the errors of the data we use as input. Thus, the extrapolation 
from $t=0$, where we fixed parameter $c$ [see Eq.~(\ref{vmdk0})], to higher 
values of $t$ is well defined. Morever, $c$ determines the relative weight 
of the $\omega$ and $\phi$ contributions, which do not differ greatly in 
our region of $t$. So, the value of $c$ is not critical.

The branching fractions are not the only quantities that can be compared to 
experimental data. The decay form factor $f(t)$, which also is very 
important, is defined through the matrix-element parametrization 
\be
{\cal M}=C\ f(t) \ \left( p_K +p_\pi\right)^\mu\ 
\overline{u}(p_-)\gamma_\mu v(p_+)\ 
\label{ffdef}
\ee
and the normalization condition $f(0)=1$. Comparing Eqs.~(\ref{myme})
and (\ref{ffdef}) reveals that the mechanism considered here
exhibits a form factor 
\be
f(t)=\frac{1}{t}\frac{\ffkp-\ffpi}{F^\prime_{K^+}(0)-
F^\prime_{\pi^+}(0)}\ .
\ee
In Fig.~2, we plot the absolute value of this form factor assuming
the VMD relations (\ref{vmdk0}) and (\ref{fk0asdiff}) with $c=0.36$ 
(method 6), and of the form factor calculated for the long-range mechanism 
dominated by the $a_1^+$ and $\rho^0$ mesons \cite{md,ratio}. 
What is finally observed experimentally is the form factor belonging 
to the superposition of those two mechanisms. This most important result 
is depicted by the solid line. Because the normalisation of the meson-dominance
matrix element is yet a little uncertain, see discussion in \cite{md}, we 
considered its multiplicative constant a free parameter. Its value was 
chosen to get the correct branching fraction for the $\die$ mode 
($2.74\dek{-7}$, \cite{pdg}). For comparison, we also show 
the linear parametrization of the form factor used by experimentalists 
\[
f(t)=1+\lambda\frac{t}{m^2_\pi}\ ,
\]
where $\lambda=0.105$ \cite{alliegro}. Taking into account the 
experimental errors of $\lambda$, which are 0.035~(stat.) and 0.015~(syst.)
\cite{alliegro}, we can show that the solid curve is compatible with the 
published experimental results. However, the preliminary data of the 
BNL-E865 collaboration \cite{hong} indicates a steeper slope, which may 
be a problem for the meson dominance model, even when it is supplemented 
with the mechanism considered in this note.

Having fixed the form factor that belongs to the superposition of 
the two mechanisms mentioned above, we can evaluate the
$\ratio$ branching ratio. It comes out as 
\[
\frac{B(\pimm)}{B(\piee)}=0.248\pm0.002.
\]
The quoted error reflects the uncertainties of the parameter $c$ and
of the $\rho^0$ running mass squared \cite{ratio} in our $t$-region. 
Using the recommended value $(2.74\pm0.23)\dek{-7}$ of the branching 
fraction of the $\die$ mode \cite{pdg}, we end up with the prediction
\[
B(\pimm)=(6.8\pm0.6)\dek{-8}.
\]

Let us compare now our formula for the matrix element and the numerical
results with those of previous studies \cite{vain,eilam}.

Our matrix element (\ref{myme}) differs from that
derived in \cite{vain}, see  Eq.~(\ref{vainformula}), even after
the form-factor difference in our formula is expressed in terms  
of the mean radii squared, using Eq.~(\ref{lowt}). We could not 
locate the source of the discrepancy. To see how serious this
discrepancy is from a pragmatic point of view, we calculated
the absolute value of the matrix element ratio using the mean
square radii of $\pi^+$ and $K^+$ of 0.44~fm$^2$ and 0.34~fm$^2$,
respectively \cite{amen1,amen2}. The result is
\[
\left|\frac{{\cal M}({\rm this~work})}{{\cal M}
({\rm Ref.~\protect\cite{vain}})}
\right|=\frac{m_K^2+m_\pi^2}{2}\left|\frac
{\langle r^2\rangle_{K^+}-\langle r^2\rangle_{\pi^+}}
{m_K^2\langle r^2\rangle_{K^+}-m_\pi^2\langle r^2\rangle_{\pi^+}}
\right|\approx 0.18\ .
\]

A formula for the matrix element of the same ``inner-bremsstrahlung" 
mechanism was proposed also in \cite{eilam}. It contains a simple 
difference of the $\pi^+$ and $K^+$ electromagnetic form factors, 
in agreement with our findings and in disagreement with (\ref{vainformula}). 
However, the same form was obtained by different means. The
authors of \cite{eilam} simply assumed that the $K^+\ra\pi^+$ transition 
amplitude was a constant, whereas in \cite{vain}, and also in our 
approach, this amplitude is proportional to the momentum squared. 
The latter is equal to the mass squared of that of the two mesons 
which is on the mass shell. In our approach
the form factor difference results from the interplay between
the momentum-dependent transition amplitude and the contributions
to the matrix element that originate in the weak-interaction part of the 
Lagrangian (\ref{l0}).

The multiplicative constant in the matrix element presented in 
\cite{eilam} is fixed, through a chain of reasoning, by the data on 
the $K\ra\pi\pi$ and $K^+\ra\mu^+\nu_\mu$ decay rates. This shows that
the physics assumed to lie behind the $K^+\ra\pi^+$ transition is different
from that here and in Ref.~\cite{vain}, where the former source of 
information is replaced by the $\pi^+\ra\mu^+\nu_\mu$ decay rate.

What concerns the magnitude of the matrix element here and in \cite{eilam}, 
their estimate 
\[
\left|\langle \pi^+|H_w|K^+\rangle\right|\approx 3.9\dek{-8}~{\rm GeV}^2 
\]
should be compared to ours
\[
\frac{G_F}{2\sqrt{2}}\left|\fpi V_{ud}\fkp V_{us}\right|
(m_K^2+m_\pi^2)=4.91\dek{-9}\ .
\]
The strikingly different magnitudes of the matrix element lead
to different judgements about the role of the considered
mechanism. In  \cite{eilam} it was deemed very important, with the
matrix element twice as big as that required by experimental branching
fraction for the $\die$ mode. Here, we found (see Table 1) that this 
mechanism, if taken alone, leads to branching fractions that
are at least one order-of-magnitude below the experimental values.

It was shown in \cite{md} that the $a_1/\rho$-meson dominance makes
the leading contribution to the matrix element of the $\pill$ decays.
Now we have shown that the matrix element of the mechanism considered here
is several times smaller (4-8 times, if we quote the results of methods 5 
and 6, which we trust the most). Neverthless, its interference
with the dominant mechanism somewhat improves the behavior
of the form factor, which is a little too flat for the dominant mechanism
alone \cite{ratio}.

\acknowledgements
I am indebted to Andrei Poblaguev for useful discussions and checking
the numerical calculations.
This work was supported by the U.S. Department of Energy 
under contract No. DOE/DE-FG02-91ER-40646 and by the Grant Agency of 
the Czech Republic under contract No. 202/98/0095.

\bt
\caption{Branching fractions of the $\piee$ and $\pimm$ decays calculated
using different inputs and compared to data \protect\cite{pdg}:
(1) $K^+$ and $\pi^+$ form factors \protect\cite{dubn}; (2) $K^0$ form factor 
\protect\cite{dubn}; (3) MRS difference of $K^+$ and $\pi^+$ 
\protect\cite{amen2}; (4) MRS of $K^0$ \protect\cite{molzon}; (5) VMD form 
factor of $K^0$ with $c$ fixed by the difference in $K^+$ and $\pi^+$ 
MRS \protect\cite{amen2};
(6) as in (5), but fixed by the $K^0$ MRS \protect\cite{molzon}.}
\label{tab1}
\btab{lccccccc}
Method & 1 & 2 & 3 & 4 & 5 & 6 & Exp. data  \\
\hline
$B(\die)\dek{9}$ & $\approx 1.3$ & $\approx 22$ & \hh{10}{11}{7} 
& \hh{2.8}{3.3}{1.0} & \hh{14}{13}{9} & \hh{4.6}{4.5}{2.9}
& $274\pm23$\\
$B(\dim)\dek{9}$ & $\approx 0.3$ & $\approx 7.4$ & \hh{1.9}{2.1}{1.3}
& \hh{0.54}{0.65}{0.40} & \hh{4.0}{3.5}{2.4} & \hh{1.5}{1.2}{0.9}
& $50\pm10$ \\
%$\ratio$ & $\approx 0.23$ & $\approx 0.34$ & 0.195 & 0.195 &
%\hh{0.292}{0.041}{0.015} & \hh{0.33}{0.06}{0.03} \\
%$\lambda_{\rm eff.}$ & $\approx 0.044$ & $\approx 0.25$ & 0 & 0 &
%\hh{0.15}{0.09}{0.03} & \hh{0.24}{0.22}{0.06} \\
\etab
\et

\begin{figure}
\begin{center}
\leavevmode
\setlength \epsfysize{15cm}
\setlength \epsfxsize{15cm}
\epsffile{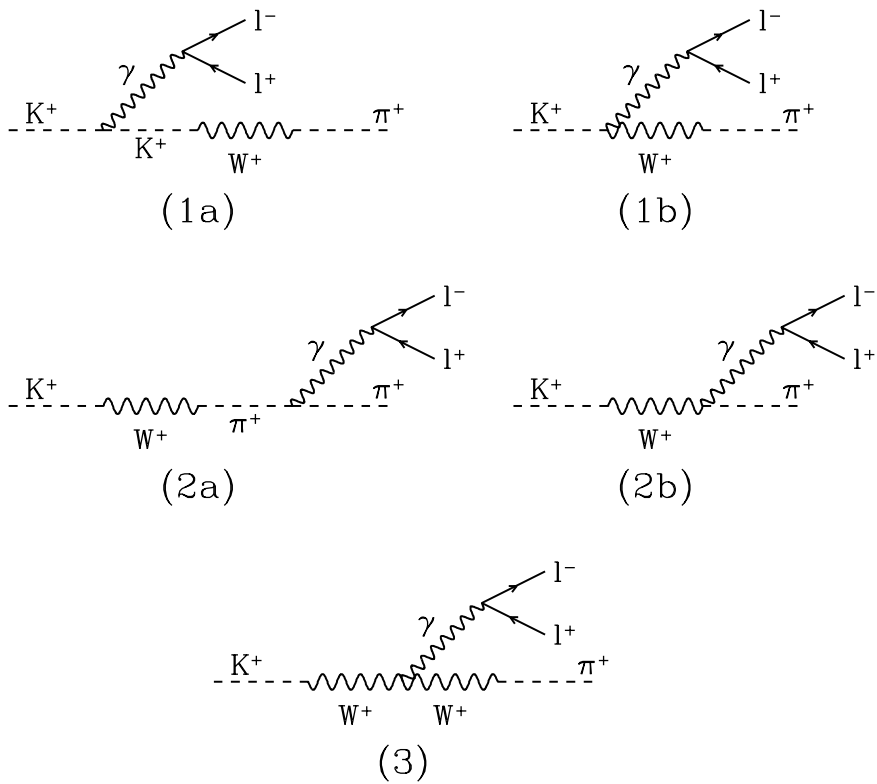}
\end{center}
\caption{Feynman diagrams contributing to the matrix element of the
\protect{$\pill$} decay. Contributions (1a) and (1b) are proportional
to the $K^+$ form factor, (2a) and (2b) to the $\pi^+$ form factor.
Contributions (1a) and (2a) were generated from the free part of the 
Lagrangian, and (1b) and (2b) from its weak-interaction part. 
Contribution (3) is suppressed by an additional factor of $(m_K/m_W)^2$ 
and therefore is ignored.
}
\label{figure1}
\end{figure}

\begin{figure}
\begin{center}
\leavevmode
\setlength \epsfysize{15cm}
\setlength \epsfxsize{15cm}
\epsffile{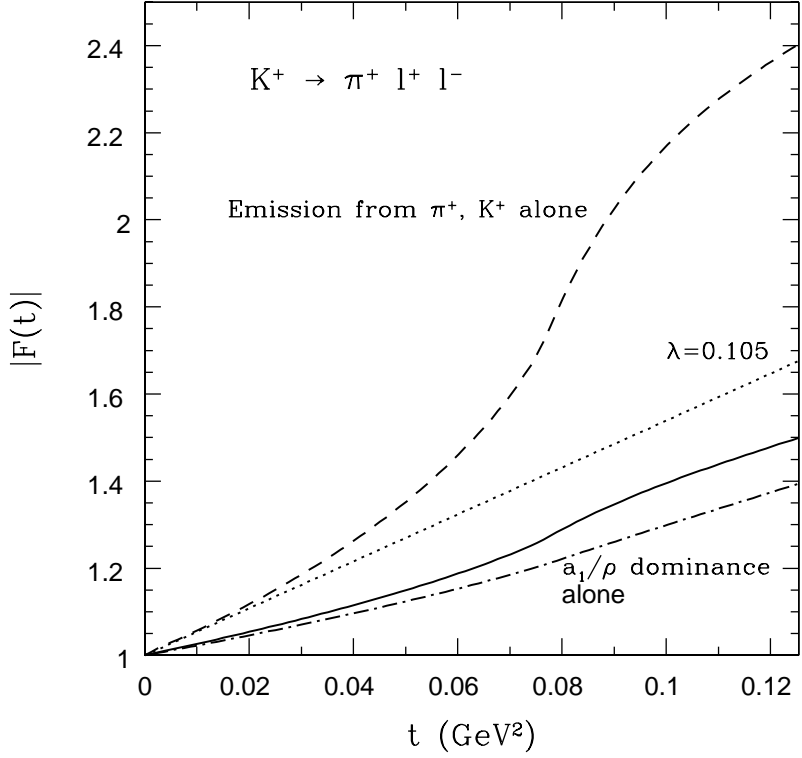}
\end{center}
\caption{Form factor of the decay $\pill$: The mechanism 
considered in this work taken alone (dashed line); 
The form factor coming from the $a_1/\rho$ dominance 
\protect\cite{md,ratio} (dash-dotted);
Superposition of the $a_1/\rho$ dominance and the present mechanism
(solid);
The linear parametrization used to fit data
\protect\cite{alliegro} (dotted).
}
\label{figure2}
\end{figure}

\end{document}